\documentclass[twocolumn,showpacs,preprintnumbers,amsmath,amssymb,prb,10pt,aps]{revtex4}
\usepackage{graphicx}
\usepackage{dcolumn}
\usepackage{bm}
\usepackage{amsmath,amsfonts,amssymb,amsthm,verbatim}
\usepackage{epsfig}
\usepackage[ansinew]{inputenc}

\newcommand{\vect}[1]{\boldsymbol{#1}}

\renewcommand{\(}{\left(}
\renewcommand{\)}{\right)}

\begin{document}
\title{Decohering the Fermi liquid: A dual approach to the Mott Transition}
 \author{David F. Mross, and T. Senthil}
\affiliation{Department of Physics, Massachusetts Institute of Technology,
Cambridge, MA 02139, USA }
\begin{abstract}
We present a theoretical approach to describing the Mott transition of electrons on a two dimensional lattice that begins with the low energy effective theory of the Fermi liquid.
The approach to the Mott transition must be characterized by the suppression of density and current fluctuations which correspond to specific shape deformations of the Fermi surface. We explore the nature of the Mott insulator and the corresponding Mott transition when these shape fluctuations of the Fermi surface are suppressed without making any a prior assumptions about other Fermi surface shape fluctuations. Building on insights from the theory of the Mott transition of bosons, we implement this suppression by identifying and condensing vortex degrees of freedom in the phase of the low energy electron operator. We show that the resulting Mott insulator is a quantum spin liquid with a spinon fermi surface coupled to a $U(1)$ gauge field which is usually described within a slave particle formulation. Our approach thus provides a coarse-grained treatment of the Mott transition and the proximate spin liquid
that is nevertheless equivalent to the standard slave particle construction. This alternate point of view provides further insight into the novel physics of the Mott transition and the spin liquid state that is potentially useful. We describe a generalization that suppresses spin anti-symmetric fluctuations of the Fermi surface that leads to a spin-gapped charge metal previously also discussed in terms of slave particle constructions.

\end{abstract}
\maketitle

\section{Introduction}

%\newpage
Describing the evolution from a metal to a Mott insulator as the strength of electron correlations is increased has challenged condensed matter theorists for the last many decades.
Part of the difficulty is due to a complicated interplay between the metal-insulator transition and magnetic ordering that appears commonly in the insulator. However in recent years evidence, both theoretical\cite{lesik,leesq,lawler,orgz2,block} and experimental\cite{kanoda0,kato,takagi,kanoda1}, has accumulated for the existence of a possible quantum spin liquid Mott state just on the insulating side of the Mott transition in frustrated lattices. Such a state does not break spin rotation or any other microscopic symmetries. Thus there is an opportunity to study the fundamental phenomenon of the Mott metal-insulator transition without the complications of magnetism or other broken symmetries.

A theoretical description of the vicinity of the Mott transition is challenged by the need to describe the kinetic and interaction energies of the electrons on more or less equal footing.
In the last two decades there has been great progress in studying the Mott transition in spatial dimension $d = \infty$ through Dynamical Mean Field Theory\cite{dmft}. However in this limit the resulting Mott insulator has at zero temperature no correlations between the spins and hence has non-zero entropy. Thus the $ d = \infty$ results may at best be a guide for the somewhat high temperature physics in the vicinity of the Mott transition.

The Mott transition can also be described using slave particle methods. A complete theory for a continuous Mott transition from a Fermi liquid metal to a quantum spin liquid Mott insulator in two space dimensions was described in Ref. \onlinecite{mottcrit} by studying fluctuations on top of the slave particle mean field theory of Ref. \onlinecite{flge}. In this description the Mott insulator is a quantum spin liquid state with gapless spin excitations. As this transition is approached from the metal the entire Fermi surface disappears {\em continuously}. The resulting critical point is characterized by the presence of a gapless `critical Fermi surface' \cite{critfs} where the Fermi surface is sharp but there are no Landau quasiparticles. These results have also been generalized to three space dimensions\cite{3dmit}.
\begin{figure}[h]
\includegraphics[width=7cm]{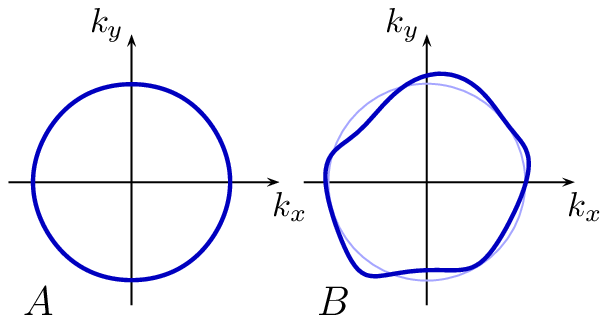}
\caption{A: Fermi surface at equilibrium. B: The low energy excitations describe smooth shape fluctuations of the Fermi surface.}
\label{fig:smooth}
\end{figure}

\begin{figure}
\includegraphics[width=8.5cm]{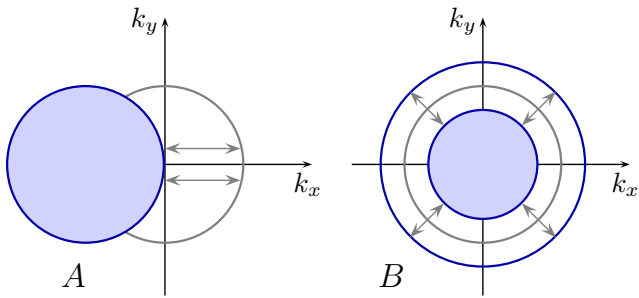}
\caption{A: Current fluctuation ($l=1$) of the Fermi surface. B: Density fluctuation ($l=0$) of the Fermi surface.}
\label{fig:breath}
\end{figure}

In this paper we develop a way of thinking about the electronic Mott transition in two space dimensions that starts with the low energy effective theory of the Fermi liquid phase. This is of course just the familiar Landau theory of Fermi liquids\cite{nozieres}. The hydrodynamic description of the low energy properties of a Fermi liquid is in terms of the kinetic equation for the {\em local} quasiparticle distribution function. This hydrodynamic theory may be usefully viewed as a theory of shape fluctuations of the Fermi surface\cite{haldane,netofr,marston,xiaogang} (see Fig. \ref{fig:smooth}). In the low energy limit a local version of Luttinger's theorem\cite{luttinger} is obeyed so that a local shape fluctuation of the Fermi surface which changes the area leads to a change of particle density (see Fig. \ref{fig:breath}). Similarly a local shift of the ``center of mass" of the Fermi surface corresponds to a non-zero current density (see Fig. \ref{fig:breath}).
As the Mott transition is approached the minimal thing that needs to happen is that long wavelength fluctuations in the local charge and current densities are suppressed. Our goal in this paper is to explore the nature of the Mott insulator and the transition when the Fermi surface shape fluctuations associated with charge and current density fluctuations are suppressed without making any a priori assumptions about the fate of other shape fluctuations.

We will do this by relying crucially on key insights from the theory of the Mott transition of bosons. In contrast to electronic systems, the Mott transition of interacting bosons is rather well understood theoretically\cite{subirbook}. The liquid phase of bosons has superfluid long range order, and the corresponding hydrodynamic theory describes the gapless sound mode associated with superfluid order.
Accessing a Mott insulator of bosons from the superfluid phase requires condensing vortex configurations of the superfluid order parameter. This enforces the quantization of particle number that is crucial for the existence of a Mott insulator.

Motivated by this, we develop an alternate `dual' way of thinking about the {\em electronic} Mott transition and the resulting spin liquid Mott insulator. We first identify a degree of freedom in the hydrodynamic description of the conducting state that plays the role of a vortex in the phase of the electron operator. We show that condensing these vortices leads to an incompressible Mott insulator. Specifically it clamps down long distance, long time fluctuations of the particle number and current densities.

 The resulting Mott insulator is in a quantum spin liquid state and has gapless excitations. Interestingly it is a spin liquid with a Fermi surface of charge neutral spin-$1/2$ fermionic spinons coupled to a fluctuating $U(1)$ gauge field.
Usually spin liquids with emergent spinons and gauge fields are described by slave particle constructions\cite{lnwrmp}. The route to a spin liquid described in this paper is very different and does not explicitly rely on any slave particle description. It thus provides an alternate point of view of the spinon Fermi surface state that may potentially be useful in thinking about it. We also briefly discuss a phase obtained by condensing vortices in a phase that is conjugate to spin antisymmetric distortions of the Fermi surface. This results in a spin gapped metal with a charge Fermi surface which can also be alternately described in terms of slave particles.

%Empirically in all known examples the quantum spin liquid state in a weak Mott insulator appear to be gapless (at least to energy scales well below any natural microscopic scale).
%Our understanding of such gapless quantum spin liquid states is rather limited. To date the only existing theoretical framework is based on fermionic slave particle representations of the underlying spins. This leads to the formulation of effective field theories of gapless spin liquids in terms of fermionic excitations coupled to a gauge field. It is therefore natural to explore alternate approaches to

\section{Overview}
\subsection{Bosons and the Mott transition}
Consider, for concreteness, the boson Hubbard model on, say, a two dimensional lattice with an integer number $N_0$ of bosons per site on average:
\begin{equation}
H_b = -t\sum_{<ij>} \left(b^\dagger_i b_j + \text{h.c.} \right) + \frac{U}{2}\sum_i \left(n_i - N_0 \right)^2
\end{equation}
where $b^\dagger_i$ creates a boson at site $i$ and $n_i = b^\dagger_i b_i$ is the boson number at site $i$. If $U \gg t$ the ground state is a Mott insulator while in the opposite limit $ t \gg U$ a superfluid state results. The superfluid-Mott phase transition is described\cite{subirbook} by a standard (quantum) Landau-Ginzburg action for the superfluid order parameter in $D = 2+1$ space-time dimensions.

For our purposes it is insightful to understand how to think about the Mott insulator and the transition to it within in a conceptual framework that begins in the superfluid phase.
Deep in the superfluid state the appropriate low energy effective theory of the superfluid phase is simply given by the harmonic (Euclidean) action
\begin{equation}
S_\text{eff} = \int d^2x d\tau \frac{\rho_s}{2}\left(\partial_\mu \phi \right)^2
\end{equation}
for the phase $\phi$ of the superfluid order parameter ($b \sim e^{i\phi}$). Here $\mu = (\vect x , \tau)$ and we have set the sound velocity to $1$. $\rho_s$ is the phase stiffness. A key point is to recognize that this ``phase-only" description has no hope of describing the Mott insulator. This is because the harmonic phase-only theory does not know about the quantization of the conjugate particle number. This quantization is crucial in obtaining the Mott insulator. Quantization of the particle number is equivalent to the condition that the boson phase be defined mod $2\pi$. Thus it is important to include vortices in the phase field and condense them to obtain a description of the Mott insulator.
As is well known\cite{halgupt,fishlee} this is conveniently described through a duality transformation where we regard the gapless linearly dispersing sound wave as a transverse photon of a $U(1)$ gauge field.
Formally
\begin{equation}
S_\text{eff} \rightarrow \int d^2x d\tau \frac{1}{2\rho_s} \left( \vect \nabla \times \vect a \right)^2
\end{equation}
Vortices in $\phi$ act as sources for the dual vector potential $a_{\mu}$, and can be easily incorporated as a boson field
that couples minimally to $a_{\mu}$. The full dual action including a vortex field $\Phi_v$ is
\begin{equation}
\label{dualboson}
S = \int d^2x d\tau \frac{1}{2\rho_s} \left( \vect \nabla \times \vect a \right)^2 + |\left(\partial_\mu - \mathrm{i} a_\mu \right) \Phi_v|^2 + V(|\Phi_v|^2)
\end{equation}
In the superfluid phase the vortices are gapped and may be integrated out to reproduce the dual of the low energy sound wave action. The Mott insulator is obtained when the vortices condense. This gaps out the gauge field, and quantizes the dual magnetic flux which is the particle number of the original bosons.
The dual vortex action is well known to be equivalent to the Landau-Ginzburg description of the Mott transition in terms of the boson order parameter $b$.

\subsection{Electrons and the Mott transition}

So now consider the electronic Hubbard model at half-filling on a non-bipartite lattice in two space dimensions:
\begin{equation}
H_e = -t\sum_{<ij>\alpha} \left(c^\dagger_{i\alpha} c_{j\alpha} + \text{h.c.} \right) + \frac{U}{2}\sum_i \left(n_i - 1 \right)^2
\end{equation}
where $c^\dagger_{i\alpha}$ ($\alpha = \uparrow, \downarrow$) creates a spinful electron at site $i$ and $n_i = \sum_\alpha c^\dagger_{i\alpha} c_{i\alpha}$ is the electron number at site $i$. For $t \gg U$ a Fermi liquid results with a sharp Fermi surface satisfying Luttinger's theorem. For $U \gg t$ on the other hand a Mott insulator obtains.
At very large $U$ the low energy physics of the Mott insulator may be described as a spin model and will often have magnetic long range order. However we will specifically be interested in the `weak' Mott insulator that obtains just on the insulating side of the Mott transition. It seems likely, for instance on a triangular lattice, that such a weak Mott insulator will be in a quantum spin liquid state with no magnetism or other broken symmetries \cite{lesik,leesq,kanoda0,kato,kanoda1}.

It is interesting, in the fermionic version of the problem, to start again with the low energy effective theory of the conducting phase and understand what needs to be done to access a Mott insulator. In the fermionic case, as the conducting phase (for $t \gg U$ ) is just the Fermi liquid, the corresponding low energy effective theory is Fermi liquid theory. Analogously to the bosonic example, the Landau Fermi liquid action by itself does not contain within it the Mott insulator that obtains with increasing $U$. We may then ask if there are special configurations (analogous to the vortices in the boson example) that have to be included and whose condensation can lead to the Mott insulator. The analogy with bosons may be sharpened somewhat by noticing that Fermi liquid theory is essentially a quadratic hydrodynamic theory for the density of fermions at each point of the Fermi surface. (For bosons the hydrodynamic mode is just the conserved total density and the quadratic phase theory just describes fluctuations of this mode. In contrast for fermions the Fermi liquid `fixed point' has emergent conservation laws corresponding to the number of fermions at each point of the Fermi surface\cite{haldane}, and the corresponding densities are emergent hydrodynamic modes that are included in the Landau Fermi liquid description). Furthermore this is equivalent\cite{haldane,netofr,marston} to a quadratic `phase' based description in terms of the phase of the fermion operator at any given point $\theta$ of the two dimensional fermi surface. This quadratic phase action is a higher dimensional generalization\cite{haldane,netofr,marston} of the familiar bosonization from one spatial dimension. However in contrast to the one dimensional example the bosonized Fermi surface theory in higher dimensions is simply a rewrite of Landau fermi liquid theory.

For the purposes of accessing a Mott insulator, it seems most interesting to explore a slightly different point of view on the Fermi liquid. As emphasized in the literature\cite{haldane,netofr}, the quadratic hydrodynamic theory for the emergent conserved densities of the $\theta$-movers
({\em i.e.} fermions associated with an angle $\theta$ on the Fermi surface) is a theory of harmonic fluctuations of the
{\em shape} of the Fermi surface (see FIG. \ref{fig:smooth}).

The fluctuations of the {\em total density} are simply determined by the fluctuations in the area of the Fermi surface (with the assumption\cite{haldane} that long wavelength fluctuations locally satisfy Luttinger's theorem). Clearly to access the Mott insulator a description where the quantization of particle number is incorporated is necessary. By analogy with bosons we may then expect that we need to include vortices in the phase variable that is conjugate to the total particle density. This conjugate phase variable is readily identified in the bosonized phase description of the Fermi liquid (as a suitable linear combination of the phase of the various $\theta$-movers). We may then anticipate that a modified theory that correctly includes vortices in this phase variable will enable description of a Mott insulator.

The rest of the paper is structured as follows: In Sec. \ref{sec.bosonization} we will briefly review bosonization of a Fermi liquid in 2+1 dimensions \cite{haldane,marston}. In Sec. \ref{sec.vortex} we show how to incorporate vortices into the phases of the fermions and thereby obtain an effective `dual' low energy theory that generalizes Fermi liquid theory. In Sec. \ref{sec.phases} we discuss the phase structure of this dual theory and show that it contains both Fermi liquid and Mott insulating phases. Furthermore we show that the Mott insulator is a quantum spin liquid that is described at low energies in terms of a spinon Fermi surface coupled to an emergent $U(1)$ gauge field, in agreement with more standard slave particle approaches. Then in Section \ref{sec.spinvortex} we show how by condensing a vortex in the phase conjugate to spin antisymmetric fluctuations of the Fermi surface, we may access a non-fermi liquid metal with a spin gap but a charge fermi surface. In Section \ref{sec.higherl} we briefly consider defining and condensing vortices in phases conjugate to shape fluctuations of the Fermi surface with non-zero angular momentum. Section. \ref{sec.summary} contains a summary of our results.

\section{Bosonization of a Fermi Liquid in 2d}\label{sec.bosonization}
Landau Fermi liquids possess a sharply defined Fermi surface where the electron momentum distribution has a jump discontinuity.
The low energy excitations may be described in terms of an electron-like quasiparticle that is well defined asymptotically close to the Fermi surface\cite{nozieres,xiaogang}.
The long distance, low energy properties of the Fermi liquid may be described in terms of a quasiparticle distribution function $n_\sigma(\vect r, \vect k, t)$ (with the spin index $\sigma = \uparrow, \downarrow$) which is well defined for $\vect k$ close to the Fermi surface, and when $\delta n_{\sigma,\vect{k}}(\vect{x},t)=n_{\sigma,\vect{k}}(\vect{x},t)-n^0_{\sigma,\vect{k}}$ is small. $n^0 = \theta(\mu - \epsilon^0_k)$ is the ground state distribution ($\mu$ is the chemical potential and $\epsilon^0_{\vect k}$ is the dispersion of a single quasiparticle). A configuration $\delta n(\vect r, \vect k)$ has the energy\small
\begin{equation}
E[\delta n] = \sum_{\vect k, \sigma} \left(\epsilon^0_{\vect k} - \mu \right) \delta n(\vect r, \vect k) + \frac{1}{2V}\sum_{\vect k, \vect k', \sigma, \sigma'} f_{\vect k, \vect k'}^{\sigma \sigma'} \delta n_{\vect k, \sigma}\delta n_{\vect k', \sigma'}
\end{equation}\normalsize
Here $V$ is the system size and $f_{\vect k, \vect k'}^{\sigma \sigma'}$ is the familiar Landau interaction function.
In the low energy limit we can ignore quasiparticle collisions and the $\delta n$ satisfy a (linearized) kinetic equation
\small
\begin{align}
\(\frac{\partial}{\partial t}+\vect{v}_\text{F}\frac{\partial}{\partial \vect{x}}\)\delta n_{\sigma,\vect{k}}=\frac{\delta(\epsilon^0_{\vect{k}}-\mu)}{V}\sum_{\vect{k}',\sigma'}f_{\vect{k},\vect{k}'}^{\sigma,\sigma'} \vect{v}_\text{F}\cdot \frac{\partial\delta n_{\sigma',\vect{k}'}}{\partial \vect{x}},\label{qpeom}
\end{align}
\normalsize
where $\vect{v}_\text{F}=\frac{\partial}{\partial \vect{k}}\epsilon^0_{\vect{k}}$ and terms of $\mathcal{O}(\delta n_{\vect{k}}^2)$ were dropped.

To address the low energy, long wavelength physics we restrict to a small band of momentum width $\Lambda_{\parallel}$ near the Fermi surface, and define
%define the density of quasiparticles moving at an angle $\theta$ with momenta restricted to a band of width $\Lambda_\parallel$ around the Fermi-surface (``$\theta$-movers'') via
\small
\begin{align}
\rho_\sigma(\theta,\vect{r},t)%&=\hspace{-6mm}\int \limits_{|\vect{k}-\vect{k}_F|<\Lambda_\parallel}\hspace{-6mm} \frac{d^2k}{(2\pi)^2} \delta n_{\sigma,\vect{k}}(\vect{x},t)\delta\(\theta-\angle (\hat{x},\vect{k})\)\nonumber\\
&\equiv \hspace{-1mm}\int_{-\Lambda_\parallel}^{+\Lambda_\parallel}\hspace{-2mm} \frac{dk }{(2\pi)^2} \delta n_{\sigma,\theta,k}(\vect{x},t).
\end{align}
\normalsize

Here $\theta$ is an angle that labels points on the Fermi surface and the corresponding Fermi momentum is $K_F(\theta)$. The integration variable $k$ is the deviation of the momentum in the direction normal to the Fermi surface at point $\theta$. Clearly we may interpret $\Delta \theta K_F(\theta) \rho_\sigma(\theta, \vect r, t)$ as the density of electrons (of spin $\sigma$) in a patch of angular width $\Delta \theta$ centered at point $\theta$ of the Fermi surface. We will dub these electrons $\theta$-movers by analogy with the familiar terms left/right movers in $d =1 $ Luttinger liquids\cite{giam}. The deviation of the total electron density $\delta \rho_\sigma(\vect r, t)$ from its mean is clearly related to the density of $\theta$-movers through
\begin{equation}
\delta \rho_\sigma(\vect r, t) = \int d\theta K_F(\theta) \rho_\sigma (\theta, \vect r, t)
\end{equation}

It will sometimes be useful to work with discrete patches: chop up the Fermi surface into $N$ patches of angular width $\Delta \theta = \frac{2\pi}{N}$ and eventually let $N \rightarrow \infty$. Then we may write
\begin{equation}
\delta \rho_\sigma(\vect r, t) = \Delta \theta \sum_i K_{Fi} \rho_i (\vect r, t)
\end{equation}

The $\theta$-mover density may be given an interpretation in terms of the fluctuation of the shape of the Fermi surface. Consider a slow long wavelength distortion of the Fermi surface where
\begin{equation}
K_{F\sigma} \rightarrow K_{F\sigma} + \delta K_{F\sigma} (\vect r, \theta, t)
\end{equation}
Assume that the Luttinger theorem relating the area of the Fermi surface to the density holds locally even for such a long wavelength disturbance of the Fermi fluid. The the change in the density is
\begin{equation}
\delta \rho_\sigma(\vect r, t) = \int d\theta \frac{K_{F\sigma}(\theta)}{4 \pi^2} \delta K_{F \sigma}(\vect r, \theta, t)
\end{equation}
The change in the density of $\theta$-movers may then be identified as
\begin{equation}
\rho(\vect r, \theta, t) = \frac{1}{4\pi^2} \delta K_{F \sigma}(\vect r, \theta, t)
\end{equation}

We may now reduce the full kinetic equation to write down the equation of motion of the $\theta$-movers. To that end we first let $\hat{v}_F, \hat{t}$ be the normal and tangential unit vectors associated with point $\theta$ on the Fermi surface. Then by integrating with respect to $k_\parallel$ we find\small
\begin{equation}
\label{eomrhotheta}
\partial_t \rho_{\sigma}(\theta) + v_F \partial_{\parallel} \rho_\sigma(\theta) + \frac{\partial_\parallel}{(2\pi)^2}\int d\theta' K_F(\theta') f_{\sigma \sigma'} (\theta, \theta') \rho_{\sigma'}(\theta') = 0
\end{equation}\normalsize
Here $\partial_\parallel = \hat{v}_F\cdot \vect \nabla$ is the derivative in the direction normal to the Fermi surface.

A bosonized phase representation is obtained by introducing for each patch $i$ a phase field $\phi_{i\sigma}$ such that
\begin{equation}
\rho_{i\sigma} = \frac{1}{2\pi} \partial_\parallel \phi_{i\sigma}\label{densityphi}
\end{equation}
Substituting in Eqn. \ref{eomrhotheta} we get an equation of motion for $\phi_{i\sigma}$ which takes the form
\begin{align}
\(\partial_t  + v_F \partial_{\parallel}\) \partial_\parallel \phi_{i\sigma} + \Delta\theta\sum_{i'} K_{Fi'} f_{\sigma \sigma'} (i, i') \partial_\parallel\partial'_\parallel \phi_{i'\sigma'} = 0
\end{align}

This is clearly a direct generalization of the familiar phase representation of the right and left moving densities in one dimensional liquids. Just as in one dimension $\phi_{i\alpha}$ will be interpreted as the phase of the $\theta$-moving fermion, as we now discuss.
The equation of motion above can be obtained from the Lagrangian
\begin{align}
L & = L_0 + L_f \\
L_0 & = \frac{\Delta \theta}{2}\sum_{i\sigma} K_{Fi} \left(-\partial_t \phi_{i\sigma} \partial_\parallel \phi_{i\sigma} - v_F\left(\partial_\parallel \phi_{i\sigma}\right)^2\right) \\
L_f & = -\frac{\left(\Delta \theta \right)^2}{4}\sum_{ii'} K_{Fi} K_{Fi'} f_{\sigma \sigma'}(i, i')\partial_\parallel \phi_{i\sigma} \partial'_\parallel \phi_{i'\sigma'}
\end{align}
We can now write the quantum partition function as an imaginary time path integral\small
\begin{align}
Z & = \int {\cal D} \phi e^{-(S_0 + S_f)} \\
S_0 & = \sum_{i\sigma} \int_{\tau, \vect r} \frac{\Delta \theta K_{Fi}}{2} \left( - i \partial_{\tau} \phi_{i\sigma} \partial_\parallel \phi_{i\sigma} + v_F\left(\partial_\parallel \phi_{i\sigma}\right)^2\right) \\
S_f & = \sum_{ii';\sigma \sigma'}\int_{\tau, \vect r} \frac{\left(\Delta \theta \right)^2}{4} K_{Fi} K_{Fi'} f_{\sigma \sigma'}(i, i')\partial_\parallel \phi_{i\sigma} \partial'_\parallel \phi_{i'\sigma'}
\end{align}\normalsize

The structure of the time derivative term in the Lagrangian determines - in an operator framework - the commutation relations for $\phi_{i\sigma}$. Of crucial importance is the following relation shown in the Appendix \ref{calc}.
\begin{equation}
\label{patchcomm}
[\phi_{i\sigma}(\vect x), K_{Fi} \Delta \theta \rho_{i\sigma}(\vect y)] = \frac{\mathrm{i}}{2\pi} D(x_\perp - y_\perp) \delta(x_\parallel - y_\parallel)
\end{equation}
Here $(x, y)_{\parallel} $ are the components of the spatial coordinates in the direction parallel to the patch normal and $(x,y)_\perp$ are the components in the direction perpendicular to the patch normal. The function $D$ is a delta-function smeared over a distance of order the inverse patch size $\sim \frac{1}{K_{Fi} \Delta \theta}$. Thus in the long wavelength limit $2\pi \phi_i$ is conjugate to the density of $\theta$-movers. We may now sum over all patches to find
\begin{equation}
[\phi_{0\sigma}(\vect x), \delta \rho_\sigma(\vect y)] = \frac{\mathrm{i}}{2\pi}\delta^{(2)}(\vect x - \vect y)
\end{equation}
with $\phi_{0\sigma}$ defined by
\begin{equation}
\phi_{0\sigma} = \frac{1}{N}\sum_i \phi_{i\sigma}
\end{equation}
Thus $2\pi \phi_{0\sigma}$ is canonically conjugate to the long wavelength fluctuations of the total density of spin $\sigma$. Defining the angular Fourier transform of the $\phi_i$ fields:
\begin{equation}
\phi_l = \frac{1}{N} \sum_i e^{-\mathrm{i}l \theta_i} \phi_i
\end{equation}
we identify
$\phi_{0\sigma}$ with the $l = 0$ component. It is thus natural that it is conjugate to the uniform ``breathing" mode of the Fermi surface which corresponds to a change of electron density. If one wishes the quadratic action for the $\phi_{i\sigma}$ fields can be readily rewritten in terms of Fourier transformed variables $\phi_{l\sigma}$.

The commutation relation Eqn. \ref{patchcomm} implies that the operator $e^{2\pi \mathrm{i}\phi_{i\sigma}}$ adds charge $1$ to patch $i$ with spin $\sigma$. In the Appendix \ref{calc} we further show that we may indeed identify this with the electron creation operator at patch $i$:
\begin{equation}
\psi^\dagger_{i\sigma} \sim e^{2\pi \mathrm{i}\phi_{i\sigma}}
\end{equation}

As in one dimensions it is now convenient to go to charge and spin bosons $\phi_{ci}, \phi_{si}$. We thus define
\begin{eqnarray}
\phi_{ci} & = & \frac{\phi_{i\uparrow} + \phi_{i\downarrow}}{2} \\
\phi_{si} & = & \frac{\phi_{i\uparrow} - \phi_{i\downarrow}}{2}
\end{eqnarray}

The action now splits into separate ones for $\phi_{ci}, \phi_{si}$:\small
\begin{align}
S & = S_{c0} + S_{cf} + S_{s0} + S_{sf} \\
S_{c0} & = \sum_{i} \int_{\tau, \vect r}\Delta \theta K_{Fi} \left( -\mathrm{i}\partial_{\tau} \phi_{ci} \partial_\parallel \phi_{ci} + v_F\left(\partial_\parallel \phi_{ci}\right)^2\right) \\
S_{cf} & = \sum_{ii'}\int_{\tau, \vect r} \frac{\left(\Delta \theta \right)^2}{4} K_{Fi} K_{Fi'} f_s(i, i')\partial_\parallel \phi_{ci} \partial'_\parallel \phi_{ci'} \\
S_{s0} & = \sum_{i} \int_{\tau, \vect r}\Delta \theta K_{Fi} \left( -\mathrm{i}\partial_{\tau} \phi_{si} \partial_\parallel \phi_{si} + v_F\left(\partial_\parallel \phi_{ci}\right)^2\right) \\
S_{sf} & = \sum_{ii'}\int_{\tau, \vect r} \frac{\left(\Delta \theta \right)^2}{4} K_{Fi} K_{Fi'} f_a(i, i')\partial_\parallel \phi_{si} \partial'_\parallel \phi_{si'}
\end{align}\normalsize
Here we defined the standard symmetric and antisymmetric combination $f_s, f_a$ of the Landau interaction function ($f_s = \frac{f_{\uparrow \uparrow} + f_{\uparrow \downarrow}}{2}, f_a = \frac{f_{\uparrow \uparrow} - f_{\uparrow \downarrow}}{2}$). The commutation relations satisfied by $\phi_{ci}, \phi_{si}$ may be readily written down. Clearly we have
\begin{equation}
[\phi_{c0}(\vect x), \delta \rho(\vect y)] = \frac{\mathrm{i}}{2\pi} \delta^{(2)}(\vect x - \vect y)
\end{equation}
with the obvious notation $\phi_{c0} = \frac{\phi_{0\uparrow} + \phi_{0\downarrow}}{2}$, $\delta \rho = \sum_\sigma \delta \rho_\sigma$.

\section{Incorporating charge vortices}\label{sec.vortex}
The Mott insulator is incompressible for density fluctuations. Thus we need to clamp down fluctuations in the shape of the Fermi surface that correspond to a change of area. Furthermore we also need to clamp down fluctuations where the center-of-mass of the Fermi surface is displaced - these correspond to a non-zero current density (see Fig. \ref{fig:breath}).

Following the logic for bosons we expect that these can be accomplished by including vortices in the phase of the fermions. Specifically if we include vortices in the phase $\phi_{0c}$ (which is conjugate to the total density) and condense them, we will get a fermionic Mott insulator.

To think about a vortex state in a Fermi liquid, first consider a ring geometry obtained by imposing periodic boundary conditions along the $x$-direction, {\em i.e.} we identify $x$ with $x + L$. Imagine slowly turning on a $2\pi$ flux. This has the effect of twisting the electron boundary conditions on going around the ring by $2\pi$. Correspondingly all single particle momentum eigenstates shift by $\frac{2\pi \hat{x}}{L}$. For a free fermi gas the ground state wavefunction is then given by
\begin{equation}
|\psi_\text{twist} \rangle = \prod_{|k_\sigma| \in FS} c^\dagger_{\vect k + \frac{2\pi \hat{x}}{L}, \uparrow}c^\dagger_{\vect k + \frac{2\pi \hat{x}}{L}, \downarrow}| 0 \rangle
\end{equation}
This shifts the entire Fermi surface by $\frac{2\pi \hat{x}}{L}$. Therefore the Fermi surface displacement is
\begin{equation}
\delta K_{F\sigma}(\theta) = \frac{2\pi}{L}\hat{v}_{F\theta}\cdot\hat{x}
\end{equation}
which can be obtained from
\begin{equation}
\phi_{ci} = \frac{x}{L}
\end{equation}

This is consistent with the expectation that the phase of the electron is shifted by $2\pi$ when a $2\pi$ twist of the boundary condition is imposed. Clearly
\begin{equation}
\oint d\vect l\cdot \vect \nabla \phi_{i\sigma} = 1
\end{equation}
for any loop around the ring. Thus there is a $2\pi$ vortex in the electron phase. This vortex is present in both $\phi_\uparrow$ and $\phi_\downarrow$. Thus $\phi_c$ winds but $\phi_s$ does not wind at all.

We may also directly consider such vortices in the bulk. Consider a ``core" such that the electron phase $\phi_{c0}$ winds by $2\pi$ on going around a loop surrounding the core.
In the presence of such a vortex $\phi_c$ will not be a smooth field anymore. We now show how the Fermi liquid action may be generalized to allow for the presence of such vortices. As in the usual discussion of the 2d XY model, we separate $\phi_{c0}$ into a smooth part $\underline{\phi}_{c0}$ and a vortex part $\phi_{c0}^v$:
\begin{equation}
\phi_{c0} = \underline{\phi}_{c0} + \phi_{c0}^v
\end{equation}
The smooth part satisfies
\begin{equation}
\epsilon_{\mu\nu\lambda} \partial_\nu \partial_\lambda \underline{\phi}_{c0} = 0
\end{equation}
The vortex part is defined in terms of the vortex 3-current $j_{\mu v}$ through the equation
\begin{equation}
\label{jvdef}
\epsilon_{\mu\nu\lambda} \partial_\nu \partial_\lambda \phi_{c0}^v = j_{\mu v}
\end{equation}

As $\partial_\mu \phi_{c0}^v$ is not really a simple gradient we will replace it by a vector field $a_\mu$ satisfying
\begin{equation}
\epsilon_{\mu \nu\lambda} \partial_\nu a_\lambda = j_{\mu v}
\end{equation}
Thus vortex degrees of freedom may be included in the Fermi liquid action by simply replacing every occurrence of $\partial_\mu \phi_{c0}$ with $\partial_\mu \underline{\phi}_{c0}
 + a_\mu$. Some care is required however in implementing this procedure. To see this consider the terms in the bosonized action involving time derivatives: it has the structure
 \begin{equation}
 \sum_{i} \int_{\tau, \vect r}\Delta \theta \left( -\mathrm{i}\partial_{\tau} \phi_{ci} \vect K_i \vect \nabla \phi_{ci} \right)
 \end{equation}
 with $\vect K_i = K_{Fi} \hat{v}_{Fi}$. In terms of the Fourier modes $\phi_{cl}$, this becomes
 \begin{equation}
 \label{stime}
 \int_{\tau, \vect r} \sum_{ll'} -i \partial_{\tau} \phi_{cl} \vect K_{-l-l'}. \vect \nabla \phi_{cl'}
 \end{equation}
 with $\vect K_l = \frac{1}{N}\sum_i e^{-il\theta_i} \vect K_i$. For smooth configurations of all the $\phi_l$ fields, this action is symmetric under interchange of $l$ and $l'$ - we simply integrate by parts twice to also interchange the time and spatial derivatives. This can be exploited to rewrite the action in several equivalent forms - for instance we may symmetrize with respect to $l$ and $l'$. However for non-smooth configurations of $\phi_l$ these different forms are not equivalent, and we need a prescription to choose between them. The physically sensible prescription which we will employ is to insist that $\phi_{0c}$ enters the time derivative term in the action in such a way as to ensure that it correctly obeys- in an operator framework- its commutation relation with the total density. To implement this, we separate terms that involve $\phi_{0c}$ from the rest to rewrite Eqn. \ref{stime} as
 \begin{equation}
 \label{stimecorr}
 \int_{\tau, \vect r} -2\mathrm{i}\partial_{\tau} \phi_{c0} \sum_{l'}\vect K_{-l'}\cdot \vect \nabla \phi_{cl'} -\mathrm{i}\sum_{l,l' \neq 0} \partial_\tau \phi_{cl}\vect K_{-l-l'}\cdot\vect \nabla \phi_{cl'}
 \end{equation}
 In the first term the sum over $l'$ does not receive any contribution from $l' = 0$ as
 \begin{eqnarray}
 \vect K_0 & = & \int \frac{d\theta}{2\pi} \vect K(\theta) \nonumber \\
 & = & \int \frac{d\theta}{2\pi} K_{F\theta}\hat{v}_{F\theta} = 0.
 \end{eqnarray}
 This follows from time reversal invariance or inversion symmetry. Then $\phi_{c0}$ enters this part of the action linearly and only through its time derivative. Therefore we can directly read off its canonical conjugate and check that it is indeed the total density. We find
 \begin{equation}
 \delta \rho = \frac{1}{\pi}\sum_{l} \vect K_l\cdot \vect \nabla \phi_{cl}
 \end{equation}
 which is readily seen to agree with our earlier expression for $\delta \rho$.
 
 As we are concerned with vortices in $\phi_{c0}$ and none in any other $\phi_{cl}, l \neq 0$, why do we not simply work with the $\phi_{ci}$ for each patch and separate out a common vortex part? The answer has to do with the ambiguity mentioned above. The same issue arises when we try to couple an external gauge field to the bosonized Fermi liquid action, and in the theory of $d = 1$ fermionic Luttinger liquids. Indeed in the presence of an external gauge field, separate conservation of the densities of $\theta$ movers is violated as the entire Fermi surface is displaced by the electric field. This is the famous anomaly\cite{anomaly1} familiar from the theory of fermions in $d = 1$. This means that we are not free to minimally couple a gauge field or to extract the vortex part independently from each $\phi_{ci}$. Rather the correct procedure is indeed to work with $\phi_{c0}$ and proceed as we did above to either introduce vortices or to couple in a gauge field. We illustrate this issue further in Appendix \ref{min} in the familiar context of  $d = 1$ Luttinger liquids.

 We may now include vortices in the bosonized action by separating out the smooth part of $\partial_\mu \phi_{c0}$ from its vortex part as discussed before. For notational convenience we shall drop the underscore from the smooth part from now on. The resulting Lagrangian density for the charge boson then takes the form\small
 \begin{align}
 \label{chargeL}
 {\cal L}_c & = {\cal L}_{ct0} + {\cal L}_{ctl} + {\cal L}_{cx} \\
 {\cal L}_{ct0} & = -2\mathrm{i}\left(\partial_{\tau} \phi_{c0} + a_{\tau} \right)\sum_{l'}\vect K_{-l'}\cdot \vect \nabla \phi_{cl'}\label{2dgauge} \\
 {\cal L}_{ctl} & = -\mathrm{i}\sum_{l,l' \neq 0} \partial_\tau \phi_{cl}\vect K_{-l-l'}\cdot\vect \nabla \phi_{cl'} \\
 {\cal L}_{cx} & = \int_{\tau, \vect r} \sum_{ll'} \left(\partial_j \phi_l + \vect a_j \delta_{l,0}\right)M^{jk}_{ll'} \left(\partial_k \phi_{l'} + \vect a_k \delta_{l',0} \right)
 \end{align}\normalsize
 Here $M^{ij}_{ll'}$ is the angular Fourier transform of the kernel in the terms in the original charge action that depend only on spatial derivatives. We note here that the vortex part $a_\mu$ couples in to the charge boson in exactly the same form as a $U(1)$ gauge field $A^\text{gauge }_\mu$, which will be important to us below. To see this it is sufficient to consider $A^\text{gauge }_\tau\neq A_{x,y}^\text{gauge }=0$, as the ambiguity discussed above (see also Appendix \ref{min}) only involves the time-derivative term in the action. The temporal component of the gauge field $A^\text{gauge }_\tau$ couples to the total charge density, \emph{i.e.}
\begin{align}
 {\cal L}_c(A^\text{gauge })={\cal L}_c\big|_{A^\text{gauge }=0} -\mathrm{i}\rho A^\text{gauge }_\tau.
\end{align}
But this is identical to (\ref{2dgauge}) for $A^\text{gauge }_\tau\rightarrow 2\pi a_\tau  $, where the $2 \pi$ reflects the identification of $2 \pi \phi_\theta$ as the phase of a fermionic $\theta$-mover. 

 Now we introduce a vortex field $\Phi_v$ whose three-current is precisely $j_{\mu v}$ (see eqn. \ref{jvdef}). Then the modified charge Lagrangian density is
 \begin{equation}
 \label{chargeL+v}
 {\cal L} = {\cal L}_c[\partial_\mu \phi_{cl} + a_\mu \delta_{l,0}] +\mathrm{i}A_\mu \epsilon_{\mu \nu \lambda}\partial_\nu a_\lambda
 +{\cal L}[\Phi_v, A_\mu]
 \end{equation}
 where $A_\mu$ is minimally coupled to $\Phi_v$. Varying with respect to $A_\mu$ then implements Eqn. \ref{jvdef}.

 We now specialize to the situation that the {\em total} charge density of electrons is such that there is one electron per site on average. Then the average flux seen by the vortices will be zero, and the vortex Lagrangian will have the structure \small
 \begin{equation}
 \label{vortexL}
 {\cal L}[\Phi_v, A_\mu] = |\left(\partial_\mu - \mathrm{i}A_\mu \right) \Phi_v|^2 + V\left( |\Phi_v|^2 \right) + \frac{1}{2e_0^2}\left(\epsilon_{\mu\nu\lambda} \partial_\nu A_\lambda\right)^2
 \end{equation}\normalsize
 where $V$ is a potential which can be expanded as a polynomial.

The resulting full charge Lagrangian described by Eqns. \ref{chargeL}, \ref{chargeL+v}, and \ref{vortexL} then taken together with the spin Lagrangian completes our modification of the basic Fermi liquid action to incorporate charge vortices.

\section{Phase structure of dual vortex theory}\label{sec.phases}
We will now use this dual vortex theory to discuss various possible phases of the fermion system.
\subsection{Fermi liquid}
When the vortices are gapped we expect to recover the Fermi liquid. To see this explicitly, we integrate out the $\Phi_v$ field from the action. In the presence of the vortex gap, this will lead to an innocuous renormalization of the coefficient of the Maxwell term for the $A_\mu$ field:
\begin{equation}
{\cal L}_\text{eff}[A_\mu] = \frac{1}{2e^2}\left(\epsilon_{\mu\nu\lambda} \partial_\nu A_\lambda\right)^2
\end{equation}
Now integrate out $A_\mu$:
\begin{equation}
 \int {\cal D} A_{\mu} e^{-{\int_{\tau, \vect r} \mathrm{i}A_\mu \epsilon_{\mu\nu\lambda}\partial_\nu a_\lambda + \frac{1}{2e^2}\left(\epsilon_{\mu\nu\lambda} \partial_\nu A_\lambda\right)^2} }
= e^{- \int_{\tau, \vect r}\frac{e^2 a_\mu^2}{2}}
\end{equation}
Thus the $a_{\mu}$ get a `mass' term and can be ignored for the low energy physics. Setting $a_\mu = 0$ we obtain just the bosonized action for the Landau Fermi liquid that we started out with.

\subsection{Mott insulator}
When the vortices condense the fluctuations of the fermion density and current will be suppressed and we will get a Mott insulator. Our formalism will enable us to obtain a `dual' description of this Mott insulator. When $\langle \Phi_v \rangle \neq 0$, there will be a Meissner effect for $A_\mu$, and it will acquire a `mass'. Integrating out $A_\mu$ will now generate a Maxwell term for $a_\mu$. Then the full effective Lagrangian in the Mott insulating phase
\begin{equation}
{\cal L}_\text{eff} = {\cal L}_c[\partial_\mu \phi_{cl} + a_\mu \delta_{l,0}] + \frac{K}{2}\left(\epsilon_{\mu\nu\lambda} \partial_\nu a_\lambda \right)^2
\end{equation}
where $K$ is a non-universal coefficient.
Rewriting this in terms of $\phi_{\uparrow}, \phi_{\downarrow}$ we get
\begin{equation}
{\cal L}_\text{eff} = {\cal L}[\partial_\mu \phi_{i \uparrow} + a_\mu \delta_{l,0}, \partial_\mu \phi_{i \uparrow} + a_\mu \delta_{l,0}] + \frac{K}{2}\left(\epsilon_{\mu\nu\lambda} \partial_\nu a_\lambda \right)^2
\end{equation}
This is the same as the Fermi liquid action except that the phases $\phi_\uparrow, \phi_\downarrow$ are minimally coupled to a $U(1)$ gauge field. This observation enables us to now go back to a fermionic representation. We introduce fermion fields $f_{\alpha i} \sim e^{2\pi \phi_{i\alpha}}$ for each patch $i$ and spin $\alpha$. The structure of the low energy effective theory for the Mott insulator is simply that of fermions at a Fermi surface coupled minimally to the fluctuating gauge field $a_\mu$. The corresponding action may be taken to be
\begin{eqnarray}
{\cal L}[f, a_\mu] & = & \bar{f_\alpha}\left( \partial_\tau - \mathrm{i}a_0 - \mu - \frac{\left(\vect \nabla - \mathrm{i}\vect a\right)^2}{2m} \right)f_\alpha \nonumber \\
& & + \frac{K}{2}\left(\epsilon_{\mu\nu\lambda} \partial_\nu a_\lambda \right)^2
\end{eqnarray}
Clearly the $f_\alpha$ field should be interpreted as a spinon, and the Mott insulator we obtain is described at low energies as a spin liquid with a spinon fermi surface that is coupled minimally to a fluctuating $U(1)$ gauge field. This is the same result as that obtained within the standard slave particle treatment of the Mott transition\cite{lnwrmp}. However we have reached it through a very different route that never involved splitting the electron into a product of slave particle operators. Thus our approach provides an alternate point of view on the spinon fermi surface spin liquid Mott insulator that adds to our understanding of it.

Let us pursue a bit more this connection between the approach we have taken in this paper and the standard slave particle description of this spin liquid Mott insulator. We see
that the chargon/holon operator of the slave particle formalism should be identified with $e^{2\pi i \phi_{0c}}$, {\em i.e.} the holon/chargon phase is essentially the $l = 0$ component of the charge boson. The bosonized version of the spinon field with spin $\sigma$ at angle $\theta$ on the Fermi surface is simply $e^{2 \pi \mathrm{i}\left(\phi_{\sigma}(\theta)
- \phi_{0c}\right)}$. Finally the magnetic flux of the emergent gauge field of the spin liquid is simply the density of the vortex field $\Phi_v$.

In the slave particle approach the internal gauge field is compact. The compactness means that ``instanton" events where the gauge flux changes by $2\pi$ are microscopically allowed. It has now been established that in two dimensional spin liquids with sufficient number of gapless spinon modes these instanton events are supressed at low energies so that the compactness can be ignored in the low energy effective theory of the spin liquid phase\cite{stableu1,sslinst}. How does this manifest itself in our approach? Since the magnetic flux of the emergent gauge field corresponds to the density of vortices, these instanton events correspond precisely to processes where the vortex decays and disappears. This is exactly what one expects in the Fermi liquid phase. As we discussed a vortex corresponds to a pattern of swirling current and in the Fermi liquid phase it will decay into a shower of particle-hole pairs. This decay process of the vortex is actually not included in our effective dual action. However as we know that the instanton events are suppressed in the spin liquid phase and right at the Mott transition we are justified a posteriori in ignoring this vortex decay.

\section{Spin-vortex condensation as a route to a spin gapped metal}\label{sec.spinvortex}
In this section we briefly digress and consider the natural question of what happens if we condense the vortex in the spin rather than in the charge boson.
Specifically let us define the zero angular momentum component of the spin boson:
\begin{equation}
\phi_{s0} = \frac{1}{N}\sum_i \phi_{si}
\end{equation}
It is easy to see that $\pi \phi_{c0}$ is conjugate to (twice) the $z$-component of the spin density $M^z = \left(\delta \rho_\uparrow - \delta \rho_\downarrow\right)$:
\begin{equation}
[\phi_{s0}(\vect x), M^z_{tot}(\vect y)] = \frac{\mathrm{i}}{2\pi}\delta^{(2)}(\vect x - \vect y)
\end{equation}
We can now contemplate phases in which fluctuations of the spin density and the spin current are clamped. This is achieved by condensing vortices in $\phi_{s0}$. Repeating the analysis of the previous section we see that the resulting paramagnet has (after refermionization) two species of fermions $d_\sigma$ at the original Fermi surface that are coupled to an emergent $U(1)$ gauge field with opposite gauge charges:
\begin{equation}
{\cal L} = \bar{d}_\sigma\left(\partial_\tau - \mathrm{i}\sigma a_0 - \frac{\left(\vect \nabla - \mathrm{i}\sigma \vect a\right)^2}{2m}\right)d_\sigma + \ldots
\end{equation}
This is an example of a spin-gapped ``algebraic charge liquid" metal\cite{acl} analogous to the holon metal that has been discussed in the context of doped Mott insulators. The coupling to the gauge field ultimate drives a pairing instability and the ground state will be superconducting. Nevertheless the spin gapped metallic ground state can be exposed if a magnetic field that suppresses superconductivity is present.

\section{Higher angular momenta vortex condensates?}\label{sec.higherl}
The theoretical aproach described in this paper to the Mott transition immediately invites the question on whether more exotic phases may be accessed by condensing vortices in higher angular momentum channels. Could we for instance introduce and condense vortices in the $d_{x^2 - y^2}$ channel of the charge boson, {\em i.e.} in $\phi_{c,2} + \phi_{c, -2}$? The field $\phi_{c2} + \phi_{c, -2}$ is conjugate to a $d_{x^2 - y^2}$ distortion of the Fermi surface. Such a vortex condensate would have rather exotic properties: it will be a compressible metal which is nevertheless ``incompressible" for such a $d_{x^2 - y^2}$ Fermi surface distortion. Furthermore the single particle spectrum will be gapped almost everywhere on the Fermi surface as adding an electron at a generic $k$-point necessarily couples in with the gapped $d_{x^2 - y^2}$ Fermi surface distortion mode.

Despite the appeal of describing such an exotic metal very simply it is not clear to us that it is legitimate to ever introduce such higher angular momentum vortices and study their condensation. The difficulty is as follows. To talk about vortices in some field it is important that it be a periodic variable. This in turn means that the physical Hilbert space is such that the conjugate variable is quantized to be an integer multiple of some basic unit. For the $l = 0$ charge or spin boson the conjugate variables are the total charge and spin densities respectively. Their spatial integrals can of course only take quantized values. Consequently $\phi_{co}, \phi_{s0}$ are periodic variables and it is legitimate to study vortex defects. For the higher-$l$ modes however, the conjugate variables are various shape distortions of the Fermi surface which in the microscopic Hilbert space are not quantized.
Consequently $\phi_{cl}$ or $\phi_{sl}$ for $l \neq 0$ should not be regarded as a periodic variable, and there is no obvious sense in defining vortices in these fields.

\section{summary}\label{sec.summary}

In this paper we described a new approach to thinking about the electronic Mott transition on a two dimensional lattice between a Fermi liquid metal and a quantum spin liquid Mott insulator. We started with the effective low energy theory of the Fermi liquid viewed as a theory of shape fluctuations of the Fermi surface. The Mott transition must involve suppression of shape fluctuations that correspond to density and current fluctuations. To implement this suppression, and based on insights from the theory of the Mott transition of bosons, we incorporated vortices in the electron phase into the bosonized description of the low energy description of a Fermi liquid. Condensing these vortices gives us a Mott insulator. We showed that the low energy effective theory of this Mott insulator has a spinon Fermi surface coupled to a fluctuating $U(1)$ gauge field. This is the same result as in the usual slave particle description of the Mott transition even though our approach is very different and did not explicitly invoke slave particles.
We also showed how we could obtain a description of a spin gapped metal with a charge Fermi surface by condensing vortices in the phase of the variable conjugate to the spin antisymmetric shape fluctuations of the Fermi surface. Such a state too has a well-known description in terms of slave particles.

When compared to slave particle theories our approach is more coarse-grained. We keep only degrees of freedom near the Fermi surface and see how to change their dynamics
so that the Fermi liquid is destroyed in favor of a Mott insulator or a spin gapped metal. Thus we keep the momentum space structure of the Fermi liquid while incorporating the minimum possible correlation effects needed to describe the Mott insulator. In contrast the slave particle approach is very much a real space construct, and momentum space structure is introduced by first doing a mean field calculation to obtain a fermi surface for slave fermions. By providing an alternate point of view we hope that our approach gives additional insight into the novel physics associated with the Mott transition and the quantum spin liquid state. In particular, our methods may provide a framework for thinking about phenomena such as the ``momentum-selective Mott transitions" \cite{kmottajm,kmottgb} suggested by various cluster extensions of DMFT, or to describe exotic phases beyond what is accessible within a slave particle framework. We leave exploration of such possibilities for the future.

We thank Tarun Grover and Brian Swingle for useful discussions. TS was supported by NSF Grant DMR-6922955. TS thanks the Aspen Center for Physics, where part of this manuscript was completed.
\appendix
\section{Some calculations from the bosonized theory}\label{calc}
For clarity we will demonstrate the calculation of the commutator $[\phi_i(\vect{x}),\phi_i(\vect{y})]$ in the non-interacting case $f(i,i')=0$. It should be clear however that the commutator is independent of $f(i,i')$ since it is determined exclusively by the time-derivative terms in the action, while $f(i,i')$ only appears with spatial derivatives of the fields. We compute
\begin{align}
&[\phi_i(\vect{x},0),\phi_i(\vect{y},0)]\\
&=\lim \limits_{\epsilon \rightarrow 0} \langle \phi(\vect{x},\epsilon),\phi(\vect{y},0)-\phi(\vect{x},0),\phi(\vect{y},\epsilon)\rangle\nonumber\\
&=\lim \limits_{\epsilon \rightarrow 0}\int_{\bar{k}_\parallel,\bar{k}_\perp,\bar{\omega}}\frac{e^{\mathrm{i}\vect{k}\cdot(\vect{x}-\vect{y})}}{\Delta\theta k_F}\frac{e^{\mathrm{i}\omega\epsilon}-e^{-\mathrm{i}\omega\epsilon}}{k_\parallel\(\mathrm{i}\omega +v_F k_\parallel\)}\nonumber\\
%&=\int d^2\bar{k}\frac{e^{\mathrm{i}\vect{k}\cdot(\vect{x}-\vect{y})}}{\Delta\theta k_F}\frac{\Theta(k_\parallel)e^{- v_F k_\parallel\epsilon}+\Theta(-k_\parallel)e^{ v_F k_\parallel\epsilon}}{k_\parallel}\nonumber\\
%&=2\mathrm{i}\int d\bar{k}_\perp \int_0^\infty d\bar{k}_\parallel\frac{e^{\mathrm{i}k_\perp\cdot(x_\perp-y_\perp)}}{\Delta\theta k_F}e^{-v_F k_\parallel\epsilon}\frac{\sin k_\parallel\cdot(x_\parallel-y_\parallel)}{k_\parallel}\nonumber\\
&=\mathrm{i}\int_{\bar{k}_\perp} \frac{e^{\mathrm{i}k_\perp\cdot(x_\perp-y_\perp)}}{\Delta\theta k_F}\lim \limits_{\epsilon \rightarrow 0}\frac{1}{\pi}\arctan \frac{x_\parallel-y_\parallel}{v_F\epsilon}\\
&=\mathrm{i}\frac{\text{sgn}(x_\parallel-y_\parallel)}{2} \frac{D(x_\perp-y_\perp)}{\Delta\theta k_F},
\end{align}
where we employ a standard point-splitting procedure and the smeared delta-function $D(x)$ is defined by the last line. Using Eq. (\ref{densityphi}) we then obtain 
\small
\begin{align}
 [\phi_i(\vect{x},0),K_F\Delta\theta \rho_i(\vect{y},0)]&=\frac{\mathrm{i}}{2\pi} D(x_\perp-y_\perp)\partial_\parallel\frac{\text{sgn}(x_\parallel-y_\parallel)}{2}\\
&=\frac{\mathrm{i}}{2\pi} \delta(x_\parallel-y_\parallel)D(x_\perp-y_\perp).
\end{align}
\normalsize
Next we are interested in the fermion correlator
\begin{align}&\langle \psi(\vect{x},\tau)\psi^\dagger(0,0)\rangle\sim\exp 4\pi^2\langle \phi(\vect{x},\tau)\phi(0,0)-\phi^2\rangle.
\end{align}
In Ref. \onlinecite{haldane} it has been shown that, in the long-wavelength limit, the fermion correlator is unaffected by a non-singular $f(i,i')$, so again we set it to zero for simplicity. It is then straightforward to compute
\begin{align}
&\langle \phi(\vect{x},\tau>0)\phi(0,0)-\phi^2\rangle\\
&=-\mathrm{i}\frac{1}{k_F\Delta \theta}\int_{\bar{\omega},\bar{k}_\parallel,\bar{k}_\perp}\frac{e^{\mathrm{i}\vect{x}\cdot\vect{k}}e^{-\mathrm{i}\tau\omega}-1}{k_\parallel}\frac{e^{\mathrm{i}\omega 0^+}}{ \(\omega -\mathrm{i}v_Fk_\parallel\)}\nonumber \\
&=\frac{2}{(2\pi)^2}\int_{k_\parallel>0} d k_\parallel\frac{e^{\mathrm{i}x_\parallel k_\parallel}e^{-v \tau k_\parallel}D(x_\perp-y_\perp)-1}{k_\parallel}\nonumber\\
&=-\frac{\ln \(L k_F\Delta \theta/\pi\)}{(2\pi)^2 } \ \ \ \ \ \ \ \ \ \ \ \ \ \text{for } \ \ \ \ x_\perp \gg \frac{1}{k_F\Delta \theta}\\
&=-\frac{ \ln \(1- \Lambda_\parallel \(\mathrm{i}x - v_F \tau\) \)}{(2\pi)^2} \ \ \ \text{for }\ \ \ \ x_\perp \ll \frac{1}{k_F\Delta \theta},
\end{align}
where $L$ is the size of the system and $\Lambda_\perp$ the momentum cut-off normal to the Fermi surface. We have now determined the fermion correlator
\begin{align}&\langle \psi(\vect{x},\tau)\psi^\dagger(0,0)\rangle\sim\exp 4\pi^2\langle \phi(\vect{x},\tau)\phi(0,0)-\phi^2\rangle\\
&= 0 \ \ \ \ \ \ \ \ \ \ \ \ \ \ \ \ \ \ \ \ \ \ \ \ \ \text{for } \ \ \ \ x_\perp \gg \frac{1}{k_F\Delta \theta}\\
&=\frac{\Lambda_\parallel^{-1}}{\Lambda_\parallel^{-1}- \(\mathrm{i}x - v_F \tau\) } \ \ \ \text{for }\ \ \ \ x_\perp \ll \frac{1}{k_F\Delta \theta},
\end{align}
where we have taken the thermodynamic limit $L\rightarrow \infty$. Via the same point-splitting procedure as above, it is easily verified that the fermion operators satisfy anti-commutation relations.
\section{Minimal coupling and vortices in bosonized action}\label{min}
In the main text we point out that a bit of care is required when coupling a gauge field to or isolating vortices from  the bosonized theory. To illustrate the issue and its resolution, here we briefly discuss the 1-dimensional case, which may be more familiar to the reader.

 The bosonized action of one dimensional spinless fermions can be written in terms of the phases $\phi_{R/L}$ of the right/left moving fermions as
\begin{align}
&\mathcal{S}= \mathcal{S}_R+\mathcal{S}_L\label{lrsep}\\
&\mathcal{S}_{R/L}= \frac{1}{4\pi }\int_{x,\tau}\mp \mathrm{i}\partial_\tau \phi_{R/L} \partial_x \phi_{R/L}+(\partial_x \phi_{R/L})^2.
\end{align}
To couple the bosons to an electromagnetic field one may be tempted to follow the minimal substitution prescription
\begin{align}
\partial_\mu\phi_{R/L}\rightarrow \partial_\mu\phi_{R/L} + A_\mu.
\end{align}
This is however incorrect as we can easily see, \emph{e.g.} by calculating the charge-density operator 
\begin{align}
\rho_\text{naive}& =\mathrm{i}\frac{\delta }{\delta A_\tau}\(\mathcal{S}_R(\partial_\mu\phi_R + A_\mu)+\mathcal{S}_L(\partial_\mu\phi_L + A_\mu)\)\\
&=\partial_x(\phi_R - \phi_L)/4 \pi,
\end{align}
while we know that the correct charge and current operators are given by
\begin{align}
\rho & = \partial_x(\phi_R - \phi_L)/2\pi \\
J & =  \partial_x(\phi_R + \phi_L)/2\pi.
\end{align}
This is not surprising. In fact, it is well known that the separation between right- and left-movers (Eq. (\ref{lrsep})) is incomplete in the presence of gauge-fields due to the chiral anomaly\cite{anomaly1}. Knowing the physical charge- and current operators we can readily write the correct action as\small
\begin{align}
 \mathcal{S}=&\mathcal{S}_R(\partial_\mu \phi_R)+\mathcal{S}_L(\partial_\mu \phi_L)+\int_{x,\tau} -\mathrm{i}\rho A_\tau   +J A_x+\frac{A_x^2}{\pi}\\
=&\mathcal{S}_R(\partial_\mu\phi_R + A_\mu)+\mathcal{S}_L(\partial_\mu\phi_L + A_\mu)\nonumber\\
&+\frac{1}{4\pi }\int_{x,\tau} \mathrm{i}(\partial_\tau \phi_{R}+A_\tau)( \partial_x \phi_{L}+A_x)-(R\leftrightarrow L).\label{correct1d1}
\end{align}
\normalsize
Note that the last term, which mixes left- and right-movers, vanishes upon integration by parts for $A_\mu=0$ and can then have an arbitrary coefficient. However for $A_\mu \neq 0$ the coefficient is uniquely given by Eq. (\ref{correct1d1}). It is only for this choice of coefficient that minimal substitution yields the correct action.\\
 Equivalenty we may rewrite the action in terms of
\begin{align}
 \varphi = \frac{\phi_+ + \phi_-}{2} \ \ \ \ \ \ \vartheta = \frac{\phi_+ - \phi_-}{2}
\end{align}
as
\begin{align}
\mathcal{S}&= \frac{1}{2\pi }\int_{x,\tau}-2\mathrm{i} \partial_\tau \varphi \partial_x \vartheta +(\partial_x \varphi)^2+(\partial_x \vartheta)^2\label{correct1d2}\\
&= \frac{1}{2\pi }\int_{x,\tau}-2\mathrm{i} (\partial_\tau \varphi+A_\tau) \partial_x \vartheta +(\partial_x \varphi+A_x)^2+(\partial_x \vartheta)^2\label{correct1d3}.
\end{align}
As above, for $A_\mu=0$, we are free to add $\int_{x,\tau}\left(\partial_\tau \varphi \partial_x \vartheta-\partial_\tau \varphi \partial_x \vartheta\right)$ to Eq. (\ref{correct1d2}) with a arbitray coefficient, but minimal substitution only yields the correct answer  (\ref{correct1d3}) if the coefficient is zero.

\end{document}